\documentclass[twocolumn,trackchanges]{aastex701}
\usepackage{tabularx}
\usepackage{caption}
 \usepackage{amsmath}

\newcommand{\okina}{\raise.5ex\hbox{`}}

\begin{document}

\title{First multi-opposition recoveries of large-orbit LSST TNOs}

\author[orcid=0009-0009-8439-509X,gname='Chantal', sname='Hemmann']{Chantal Hemmann}
\affiliation{Department of Physics and Astronomy, University of British Columbia, 6224 Agricultural Road, Vancouver BC, Canada}
\email[show]{chantalhemmann@gmail.com}  

\author[orcid=0000-0002-0283-2260,gname='Brett', sname='Gladman']{Brett Gladman} 
\affiliation{Department of Physics and Astronomy, University of British Columbia, 6224 Agricultural Road, Vancouver BC, Canada}
\email[show]{gladman@astro.ubc.ca}

\author[orcid=0000-0002-0283-0000,gname='Justine', sname='Obidowski']{Justine Obidowski} 
\affiliation{Department of Physics and Astronomy, University of British Columbia, 6224 Agricultural Road, Vancouver BC, Canada}
\email{jusdowski@phas.ubc.ca}

\author{Jean-Marc Petit} 
\affiliation{Institut UTINAM UMR6213, CNRS, Univ.~Marie et Louis Pasteur, OSU Theta F-25000 Besançon, France}
\email{Jean-Marc.Petit@normalesup.org}

\author{JJ Kavelaars} 
\affiliation{Herzberg Astronomy and Astrophysics Research Centre, National Research Council of Canada, 5071 West Saanich Rd, Victoria, British Columbia V9E 2E7, Canada}
\affiliation{Department of Physics and Astronomy, University of British Columbia, 6224 Agricultural Road, Vancouver BC, Canada}
\email{jj.kavelaars@nrc-cnrc.gc.ca}



\begin{abstract}


We report 2026 CFHT second-opposition recovery observations of three large-orbit Trans-Neptunian Objects (TNOs) discovered by the Vera C. Rubin Telescope's Legacy Survey of Space and Time (LSST) during preliminary observations in 2025.
Due to their potential scientific interest, we selected three TNOs whose nominal semimajor axis could be larger than 100~au according to their short-arc ($\leq0.2$-yr) LSST observations.
These three targets (2025 LS2, 2025 ME278, 2025 MX348) had discovery opposition orbits of varying observation quality (in terms of the duration and distribution of observations).
In two cases, the 2026 CFHT recoveries showed that the semimajor axis dropped by a factor of $\simeq$3--10, and that the object appears to be a (common) plutino in the 3:2 mean-motion resonance with Neptune.
We discuss how and why the estimated orbital elements evolved, and re-caution the community to be skeptical of unusual TNO orbits until they have multi-opposition recoveries.

\end{abstract}



\section{Introduction}

Determining the detailed orbital structure of the Trans-Neptunian Object (TNO) region for perihelion distances $q>50$~au or for orbital inclinations $i>50^\circ$ is 
the current holy grail of outer Solar System science
\citep{2021ARA&A..59..203G}.
Only a handful of these objects are known because they are extremely rare on
the sky ($\sim 0.1$\% of known objects).
With an estimated yield of $\sim$30,000 new TNOs \citep[][]{2025AJ....170...99K}
the Vera C.~Rubin telescope's Legacy Survey of Space and Time (LSST) offers
the possibility of increasing this sub-sample by up to an order of magnitude,
after a critical period of precise orbital determination.

Although the heavily-eroded scattering population \citep[][]{2008ApJ...687..714V}
includes semimajor axes out to $a>$1000~au  \citep{2019AJ....158...43K}, these planet-coupled TNOs largely keep $q$ below the (fuzzy)  boundary 
of $\simeq37$~au due to gravitational interactions with the giant planets \citep{2002Icar..157..269G,2007Icar..189..213L}.
The dynamics of mean-motion resonances can mildly raise a TNO's $q$, but such objects will tend to remain in or near the resonance.  
If some other physical process lifts $q$ to higher values (e.g., resonant drop-outs during either planetary migration \citep[][]{2005CeMDA..91..109G}
or due to now-absent rogue planets modifying TNO semimajor axes \citep[][]{2022ApJ...938L..23H}), 
the TNO becomes `detached' from the strong planetary scattering regime \citep[][]{2008ssbn.book...43G}.
\cite{2023PSJ.....4..145B} shows that the detached population has the expected steady-state semimajor axis power law that would arise if the objects were initially dispersed by giant-planet scattering \citep[][]{2026AJ....171..133H}.

Most extreme are the even larger-$q$ subset of the detached population informally known as the `sednoids'.
Despite the lack of a clear dynamical definition of this sub-population, the consensus within the community is that they are objects with perihelia sufficiently high such that some even more dramatic process(es) (than those which created the lower-$q$ detached TNOs) operated.
To date, there are only 4 known sednoids:
the namesake Sedna \citep[][]{2004ApJ...617..645B},
2012 VP113 \citep[][]{2014Natur.507..471T},
Leleākūhonua \citep[][]{2019AJ....157..139S}, and
Ammonite \citep[][]{2025NatAs...9.1309C}.
Having large $a$ and $e$, sednoids are only bright enough to be detected while in the small fraction of their orbits near perihelion.

The LSST began survey operations in July 2026.  
Only a tiny fraction of its detections will be sednoids or $i>50^\circ$, as they are extremely rare at the 24th magnitude flux limit ($\sim10^{-3}$ of TNO detections).
An intrinsic problem with their identification is that it is notoriously difficult to determine the $a, e, q$ elements of objects discovered at large distances with only single-opposition observational arcs.
In fact, the strong majority of objects that are detected at heliocentric distances $d>50$~au \citep[][]{2016AJ....152..221S}
will be scattering, resonant, or detached TNOs with $q$ significantly smaller than their detection distance.
It takes an abundance of observation (in number, total time base, and a variety of elongations relative to opposition) before the fit to a large semimajor axis orbit will reliably converge.
Because of the huge scientific interest, rapid identification of the large-$a$ sample and accurate $q$ determination is needed.
We here describe three cases of this process for TNOs discovered in `early' LSST data obtained in 2025 \citep[][]{2025epsc.conf..242J}, which was {\it not} acquired with the same cadence as the coming survey.
   
   
\begin{deluxetable*}{cccccccccc}
\digitalasset
\tablewidth{\textwidth}
\tablecaption{Osculating orbital elements before and after follow-up observations \label{Table1}}
\tablehead{
\colhead{Orbit} & \colhead{Arc (yr)} & \colhead{d (au)} &\colhead{q (au)} & \colhead{a (au)} & \colhead{e} & \colhead{i ($^\text{o}$)} & \colhead{node ($^\text{o}$)} & \colhead{$\omega$ ($^\text{o}$)} & \colhead{RMS(\arcsec)}
}
\startdata
\multicolumn{10}{c}{\textit{\textbf{2025 LS2}}} \\
MPC & $0.200$ & $44.90\pm0.08$ & $43 \pm 6$ & $540 \pm 1400$ & $0.92 \pm 0.21$ & $12.6 \pm 0.3$ & $144.10\pm0.3$ & $190\pm40$ & $0.12$ \\
BK & $0.200$ &  $45.00 \pm 0.08$ & $\approx43.7$ & $440 \pm 170$ & $0.90 \pm 0.04$ & $12.60 \pm 0.08$ & $144.10\pm0.07$ & $189\pm10$ & $0.13$\\
\hline
BK & $0.960$ & $45.00\pm 0.02$ & $\simeq43.3$ & $510\pm110$ & $0.91 \pm 0.02$ & $12.60 \pm 0.01$ & $144.10\pm0.01$ & $191\pm4$ & $0.13$\\ 
MPC & $0.960$ & $44.75\pm 0.07$ & $43\pm2$ & $530\pm440$ & $0.92 \pm 0.07$ & $12.62 \pm 0.03$ & $144.09\pm0.03$ & $192\pm12$ & $0.12$\\ 
\multicolumn{10}{c}{} \\
\multicolumn{10}{c}{\textit{\textbf{2025 ME278}}} \\
MPC & $0.066$ & $32\pm4$ & $17\pm60$ & $134\pm5330$ & $0.9\pm1$ & $12\pm3$ & $216\pm11$ & $307\pm60$ & $0.07$ \\
BK & $0.066$ & $31.4\pm0.5$ & $\approx29.7$ & $36\pm7$ & $0.17\pm0.82$ & $11.3\pm0.6$ & $214\pm 2$ & $350\pm280$ & $0.09$ \\ 
\hline
BK & $0.910$ & $31.32 \pm 0.01$ & $\simeq27.4$ & $39.3\pm0.3$ & $0.30\pm0.01$ & $11.085\pm0.002$ & $213.64\pm0.01$ & $101\pm1$ & $0.09$ \\
MPC & $0.910$ & $31.04\pm0.03$ & $27.4\pm0.9$  & $39\pm1$ & $0.30\pm0.04$ & $11.094\pm0.07$ & $213.67\pm0.03$ & $101\pm4$ & $0.06$\\
\multicolumn{10}{c}{} \\
\multicolumn{10}{c}{\textit{\textbf{2025 MX348 = 2015 KF176}}} \\
MPC & $0.082$ & $35\pm4$ & $18\pm61$ & $515\pm83500$ & $0.97\pm1$ & $1.8\pm0.4$ & $290\pm12$ & $230\pm30$ & $0.14$ \\
BK & $0.082$ & $34.0\pm0.2$ & $\approx27.9$ & $41\pm16$ & $0.3\pm0.6$ & $1.78\pm0.04$ & $292\pm 1$ & $240\pm40$ & $0.23$ \\
\hline
BK & $0.992$ & $33.83\pm0.01$ & $\simeq32.4$ & $39.25\pm0.14$ & $0.15\pm0.01$ & $1.752\pm0.001$ & $293.08\pm0.04$ & $340\pm10$ & $0.20$ \\ 
\hline
BK & $11.092$ & $33.827\pm0.003$ & $\simeq33.43$ & $39.23\pm0.02$ & $0.1479\pm0.0004$ & $1.7512\pm0.0005$ & $293.090\pm0.004$ & $338.35\pm0.03$ & $0.14$\\
MPC & $11.092$ & $\simeq33.821$ & $33.43\pm0.01$ & $39.10\pm0.07$ & $0.145\pm0.001$ & $1.7560\pm0.0004$ & $292.96\pm0.01$ & $338.3\pm0.1$ & $0.13$\\
\enddata
\tablecomments{The evolution of orbits using BK \citep{2000AJ....120.3323B} 
and MPC (IAU Minor Planet Center) orbit fitting as more information was
acquired. 
Solid lines separate data sets of different arc lengths.
The MPC orbits are heliocentric while those from BK are barycentric;  $d$ values are 
at the time of the first LSST astrometric point (in the respective heliocentric or barycentric frame).
We reproduce here the MPC website estimates for uncertainties and 'normalized' RMS relative to the fit, on or before 1 July 2026.}
\end{deluxetable*}
   

\section{Observations}

We have target-of-opportunity programs approved on the Canada-France-Hawaii 3.5m Telescope (CFHT) during the 2026A and 2026B semesters, designed to provide rapid orbit improvement for a moderate number of the most potentially interesting TNOs discovered by LSST.
Some of this time allocation is in `snapshot mode', designed for observations expected to yield scientific value even in poor image quality (IQ $>$1.1\arcsec) and potentially non-photometric conditions.

The typical moving-target depth for LSST\footnote{
As Rubin's systems are tuned, performance should improve to reach the $m_r\sim24.0$ design target for moving sources}
in 2025 was $m_r\simeq23.5$ \citep{2026ApJ...996L..33G}.  
Although smaller in aperture than the 8-m class Rubin obsevratory, CFHT easily recovers any Rubin detection made during normal LSST operations even during poor seeing (1.1\arcsec-1.7\arcsec) due to CFHT's ability to use longer exposures---for example, an 8-minute CFHT exposure in seeing twice as bad as Rubin's will reach the same depth
as LSST's 30-s images.
Because our short recovery sequences can `fit into holes' in CFHT queue observing operations, we received some excellent imaging even in snapshot mode (a few of the 
2025 ME278 observations described below have IQ=0.5\arcsec).
We additionally have some high-priority time in semester 2026B reserved for the most urgent recoveries.

\subsection{Target selection}

We monitor the electronic Daily Orbit Updates published by the IAU Minor Planet Center (MPC) for $q>7.35$~au outer Solar System objects
\citep[][]{2008ssbn.book...43G}
that are either 1-opposition orbits or new 2-opposition orbits that have recently been recovered or updated.
In 2025, this set was in majority LSST detections, which we expect will completely dominate the detections as normal operations commence.

Before every CFHT dark run in which the MegaPrime CCD mosaic is mounted, we assess the available targets.
For highly uncertain initial recoveries, the large ($\pm37$\arcmin) field of view 
of Megaprime is critical.
Based on decades of TNO-tracking experience, we select targets that we believe have a chance of actually having a large semimajor axis ($a>100$au), preferably with perihelion distances of $q>40$~au\footnote{At times we may select $i>50^\circ$ objects with much lower $q$ as these TNOs are also of interest.}.
This is a non-trivial task, as the short-arc orbits from the discovery opposition tend to have order unity uncertainty in $a$, $e$, and $q$---the latter is especially true when a TNO with discovery distance $d>40$~au has an unknown $e$ (and thus $q$ could be (i) interior to Neptune's orbit, (ii) at a more `normal' 32--37~au range, or (iii) actually have a higher (detached) pericenter distance).
From these potentially interesting candidates, we select roughly half a dozen targets to program into the CFHT Queued Service Observing (QSO) system.

The target's rate of motion during a particular dark run influences observation planning, as it determines the cadence required for its recovery.
If the TNO is moving fast enough ($\ga$1\arcsec/hr) that a frame sequence in a single 
night would yield detectable object motion during the above-horizon time that the target is available,
the sequence was programmed as a 4- or 6-image dither pattern to avoid CCD defects and fill in CCD gaps, while being long enough to detect the TNO's motion.
Alternatively, if the motion is too slow for this, only one observation per night is used, and one or more follow-up images on a different night are used to detect motion.
This was often programmed using the CFHT QSO REEL relative-timing feature 
(which triggers future exposures after an initial observation is acquired, on a timescale of hours to days).
In such cases, a single 8-min exposure is sufficient to reach the desired depth, and is a short enough that it is easier to flexibly schedule in between other programs, thus increasing the likelihood a second image is captured later on during the dark run.

During the March to June 2026 period we recovered three potentially large-$a$ TNOs.
These objects were all initially fit with $a>$100~au orbits by the MPC, albeit with stated semimajor axis uncertainties larger than the nominal $a$ (that is, they have order unity errors in $a$, and thus also in eccentricity).
We note that for these short-arc orbits, the JPL Horizons system essentially uses the MPC orbit (D. Farnocchia, private communication 2026) and thus does not provide an independent orbit fit; we therefore do not report the JPL estimates here.
We also computed another fit using the \citet{2000AJ....120.3323B}
code (abbreviated BK hereafter), which provides best-fit barycentric orbital
elements as opposed to the heliocentric orbits from the MPC and JPL\footnote{For
$a$=100--500~au orbits, the difference between the heliocentric and barycentric
semimajor axes can be tens of au.}
For TNO with arcs of about 30 days or less, the BK algorithm tends to
restrict large $a$ and $e$ solutions to avoid over-fitting a small
amount of data with a full 6-dimensional parameter search; in the great
majority of cases the resulting lower-$e$ orbit will turn out to be closer 
to reality for TNOs \citep[see][]{2010AJ....139.2249J}.

The images for each object were aligned and human operators visually identified a target moving with the predicted motion.
A plate solution was carefully constructed using a custom IRAF script and the 
Gaia DR3 astrometric catalogue \citep[][]{2016A&A...595A...1P,2023A&A...674A...1G}.
Depending on a target's location on a CCD and the catalogue's star density, 
we used solutions that were either very local (linear) or a higher-order polynomial to account for the change in plate scale across the CCD.
The recovery of each object was confirmed by a typically 0.1-arcsecond residual new orbital solution over our multiple 2026 nights.
In all cases reported here, the centroiding error on the TNO
itself dominates astrometric uncertainty due to their faintness.
Our astrometry (already available on the MPC) includes image-by-image uncertainty estimates that take into account the TNO's SNR, the plate solution scatter, and any crowding effects.
Initial and updated orbits for all three objects are reported in Table \ref{Table1}.

\section{Results}
\subsection{2025 LS2} 

This $m_r\simeq22.9$ LSST TNO detection had an unusually long measurement suite for LSST's 2025 season, having been observed on 11 different nights over 70 days.
Both the MPC and BK initial orbit fits agreed on the detection distance and perihelion distance (about 45 and 43 au respectively), the $i\simeq12.6^\circ$ orbital inclination,
and the node and pericenter angles.
An $a\gg$100~au was felt to be required (A.~Heinze, J.~Kurlander, and M.~Holman, 2026, private communication).
The uncertainty on the perihelion $q$ may have allowed it to rise
above 50~au, so this TNO was targeted for recovery.

The slow rate of motion of 2025 LS2 in May 2026 (well before opposition) required the recovery to be programmed as one measurement per night over two nights. 
The recovery consisted of a single 8-minute exposure acquired on May 20 2026 UT and a subsequent image on May 21 2026 UT.
The on-sky BK uncertainty ellipse was a relatively small 20$\arcsec$
due to the multi-month 2025 arc; the recovery was 7$\arcsec$ E of the predicted BK position (that is, within its estimated uncertainty ellipse) but 82$\arcsec$ E of MPC-predicted position.
Due to the sparser reference star density and a recovery in a CCD
center, a 4th-degree polynomial plate solution was used, giving a plate
solution with 
scatter $\simeq$0.03\arcsec.

The revised BK best-fit orbit based on this recovery confirmed the previously very uncertain semimajor axis to be $\simeq$500~au; the barycentric semimajor axis rose 70~au.
The barycentric elements are much preferred for large-$a$ orbits because a heliocentric semimajor axis is subject to strong short-term variations (the planets change the solar velocity on timescales less than the TNO's orbital period).
As is usual for large orbits discovered near perihelion, while the perihelion distance is relatively well determined, the $a$ and $e$ remain poorly constrained and an uncertainty $\delta(a)>10$~au will likely persist for several years \footnote{How quickly this converges will depend on a complicated fashion on the observation cadence, total arc length, astrometric uncertainty, observation elongation from observation, heliocentric distance, and true semimajor axis.}.
The angular orbital elements underwent no significant changes  following this orbit refinement (with uncertainties dropping a factor of a few).
This recovery produced the first published IAU Minor Planet Electronic Circular (MPEC 2026-N55) for a Rubin TNO discovery, due to its confirmed large $a$ (see
MPEC 2025-R47 for the MPC policy on TNO MPECs).

\subsection{2025 ME278}

Based on a 24-day arc, this $m_r\simeq22.7$ TNO was discovered at a heliocentric
distance $d\simeq31$~au (with the BK uncertainty being $<1$~au).
The MPC 1-opposition orbit placed the TNO near $d=32$~au, with estimated $(a,e)$=(134 au, 0.9) and pericenter distance $q$=17~au, classifying it as a scattering orbit (which would
make the orbit short-lived); Table \ref{Table1} indicates all of $q,a,e$ were assigned
order unity uncertainties.
In contrast, the BK code chose to avoid large-$e$ orbits given the short 24-day 
arc in 2025 \citep[as described in][]{2000AJ....120.3323B};
this resulted in a lower-$a$ orbit with perihelion slightly smaller
than the discovery distance.

Due to the rapid rate allowing single-night motion detection,
2025 ME278 was first recovered on a 6-image sequence spanning 32 minutes on 2026 May 14 UT. 
Despite being taken in snapshot mode, this suite was acquired in IQ$\simeq$0.5\arcsec conditions, between longer duration queue-mode programs. 

For the May 2026 recovery, the BK code estimated an on-sky uncertainty 
ellipse with 1-$\sigma$ major axis $\simeq$1000$\arcsec$ long.
This large uncertainty is due to the fact that the 2025 observations spanned
one month (equivalent to two sequential dark runs) and an enormous range of
semimajor axes---out to several thousand au---can satisfactorily fit
the observations.
The recovered position was in fact 293$\arcsec$ east of the BK prediction 
(that is, within 1$\sigma$) 958$\arcsec$ east of the MPC prediction.
Luckily the CFHT Megaprime field of view covered (barely) the entire 2-$\sigma$
error ellipse.

The May 2026 recovery indicated such a different orbit that other CFHT observations were needed; luckily the queue had acquired scattered frames in the two previous dark runs.
Using the May recovery, predicted the position to much higher precision and identified the target in 3 (of 6) lower-quality CFHT images we acquired in April 2026 -- the others were contaminated by excessive sky noise, a cosmic ray, or had low SNR for the TNO.
The resulting orbit refinement lowered 2025 ME278's on-sky uncertainty to  $<$1\arcsec, which made it identifiable in several poor-quality exposures on March 25, 2026;
only one of these measures was reported.
Section~\ref{sec:Discussion} discusses the dramatic orbital revision to what appears to be a plutino 
(an object in the 3:2 mean-motion resonance with Neptune).

\subsection{2025 MX348}

Based on the 30-day arc of the 2025 observations of this $m_r\simeq22.9$ TNO,  
2025 MX348's $d$ and $q$ were estimated by the MPC to be very similar 
to the initial orbit of 2025 ME278, with a large $a$ and $e$ and resulting
$q$ interior to the orbit of Uranus, but with order unity errors on these
quantities (Table \ref{Table1}).
The BK fit again proposed a more circular orbit with $q$ just inside 
Neptune's orbit, but with larger residuals (see Discussion).

During the June 2026 CFHT dark run, 2025 MX348 was moving quickly enough to be recovered 
via a 4-image dither pattern covering 24 minutes on June 13 2026 UT.
The BK code predicted an on-sky uncertainty spanning a $\pm180\arcsec \;$ 
 1--$\sigma$ ellipse.
The TNO was recovered 247\arcsec west of the BK prediction, and 594\arcsec west of the MPC prediction; 
the size of these discrepancies (relative to the uncertainty) 
indicating that the true $a$ and $e$ are different than both initial estimates.

Because the TNO was near a CCD edge, a locally-linear plate solution was constructed, having a stellar scatter of 0.02\arcsec.
During the sequence, 2025 MX348 steadily approached a background source.
The third image is much poorer quality and was excluded from the reported astrometry, while confusion with the background source produced a higher (0.2\arcsec) centroid uncertainty for the fourth image.

With the one-year arc provided by the June 13 2026 data, $a$ dropped by more than an order of magnitude from the 
1-opposition MPC orbit, with the BK fit now suggesting an even more circular orbit, with $a\simeq$39.5~au.
This first single-night recovery thus suggested that the TNO is also a plutino,
but the $a$ uncertainty was larger than the resonance width of $\simeq0.4$~au.
Therefore, another single image (with a high SNR$\approx30$) was acquired on 2026 June 21 UT to refine the orbit (Table \ref{Table1}) to one with $\delta(a)$=0.14~au, which confirmed the plutino identification (which is unremarkable except for the unusually low $i$).

Upon MPC submission of our CFHT recovery, the MPC linked 2025 MX348 back to a previously-designated object 2015 KF176; that designation was based on three observations from 2015 May 19, 20, and 23 UT that is probably associated with 
the Dark Energy Camera (DECam) survey for Near-Earth Objects described in \cite{2017AJ....154..170T}.
We have been unable to ascertain the orbit initially assigned to 2015 KF176, but with the 2026 
recovery the orbit was now accurate enough to establish the linkage back to the 2015
observations, and the MPC has assigned discovery credit to the DECam observations.
The orbit is now very accurate (with the differences between the 11-year BK and MPC
orbits being entirely due to the latter being heliocentric).

\section{Discussion}
\label{sec:Discussion}

These orbital progressions follow a well-known pattern
\citep{2010AJ....139.2249J}.
The orbit determination problem is fundamentally degenerate: for short-arc  TNOs, even excellent astrometry (like that provided by LSST) cannot get around the fact that the astrometry can be represented equally well (to better than the measurement uncertainties) by a large range of heliocentric orbits.
An unconstrained search over the 6 dimensions of orbital elements will almost always produce an orbit with smaller residuals by allowing $a$ and $e$ to become very large, and $\omega$ to take on any value---essentially, over-fitting the short-arc data.
This effect is illustrated in Table \ref{Table1} for all 3 TNOs: the unconstrained MPC fits have lower RMS residuals in 2025 relative to the observations, while the BK code, which resists this over-fitting, ended up predicting orbits closer to reality for the two plutinos based on the same 2025 data.
For 2025 LS2, the initial MPC orbit was closer for the (still rather uncertain) semimajor axis, but the BK prediction was again much closer to the recovery.

We do not believe that any particular orbit fitting approach is consistently superior.
The problem's innate degeneracy means that the "best-fit" orbit depends heavily on the assumptions/priors 
that are made.
Unless several months of arc are available in the discovery opposition (as was the case for 2025 LS2), proposed large-$a$ orbits can be wildly incorrect and should not be trusted until confirmed on a second opposition.
We note the very significant argument of perihelion ($\omega$) changes  for both 2025 ME278 and 
2025 MX348, which changed by $\approx100^\circ$; thus the perihelion longitudes should not be used 
for dynamical studies.


As it begins normal operations, LSST is expected to re-recover these targets in 2026.
Our work is complementary to this process in a number of ways:

\noindent
1. Our work is often done at large solar elongations, as the TNO
disappears behind the Sun in the discovery opposition (which we are doing from some objects discovered in 2026) or as it re-emerges the following year.
The resulting larger heliocentric parallax is of great value to the long-term orbital determination, especially for large-$a$ orbits. In addition, for the fainter TNOs, the best-IQ CFHT recoveries will have higher SNR and more accurate astrometry than the typical 30-s LSST frame.

\noindent
2. With on-sky uncertainties of $\sim$10--1000\arcsec, the MPC orbit will not at first automatically be identified in each new frame; instead, the survey will essentially `re-discover' them over some number of weeks/months during the 2026 opposition, and this `new' detection will eventually be linked to the previous designation.
With a single early recovery reported to the MPC from our program, the updated orbit will match the LSST night-by-night detections and thus immediately update the MPC orbit, circumventing the rediscovery process entirely.
This would allow for earlier follow-up study on the most interesting TNOs, which may otherwise have to wait until the 3rd opposition.

\noindent
3. A hopefully rare scenario is that extended poor weather\footnote{Which is unfortunately currently occurring in August 2026} or engineering issues could result in a specific target not being recovered at all 
in the 2nd-opposition data.
For these most interesting TNOs, tracking them with other facilities will allow the forefront science enabled by their existence to begin.

In summary, despite high-quality astrometry, 1-opposition orbits for putatively large-$a$ TNOs have significant ($\geq 100$~au) uncertainties on their semimajor axes.
This is not due to insufficient orbit fitting algorithms; only additional astrometry can defeat the degenerate nature of the problem to allow the accurate identification of the small fraction of detected TNOs which are, in fact, the most dynamically interesting.

\begin{acknowledgments}
We acknowledge helpful discussions with Matthew Holman, Ari Heinze, and Jake
Kurlander regarding the release timing of the LSST-detected TNOs, and for pointing
out 2025 LS2 as a specific target.
We thank Wesley Fraser for a careful reading which improved the manuscript.
BG and CH thank NSERC for Canadian financial support.  
We thank the 
CFHT QSO operations for their assistance in obtaining these
observations.

Based on observations using the Canada-France-Hawai\okina{}i Telescope (CFHT), operated by the National Research Council of Canada, the Institut National des Sciences de l'Univers of the Centre National de la Recherche Scientifique of France, and the University of Hawai\okina{}i, obtained with MegaPrime/MegaCam (a joint project of CFHT and CEA/DAPNIA).
Maunakea is a sacred site to Native Hawaiians, also known as Kānaka \okina{}Ōiwi. 

This work has made use of data from the European Space Agency (ESA) mission
{\it Gaia} (\url{https://www.cosmos.esa.int/gaia}), processed by the {\it Gaia}
Data Processing and Analysis Consortium (DPAC,
\url{https://www.cosmos.esa.int/web/gaia/dpac/consortium}). 

\end{acknowledgments}

\begin{contribution}

BG, JO, JK, and JP wrote the observing proposals that 
acquired the observing time.
CH wrote the MPC DOU monitoring system used to pick out the
most interesting targtes.
CH, BG, and JO programmed the CFHT observations.


\end{contribution}

%
\facilities{CFHT, Vera C. Rubin}

\software{IRAF}
\bibliography{ADS_CH_ApJL_2026}{}
\bibliographystyle{aasjournalv7}



\end{document}